\documentclass{article}
\usepackage[preprint]{log_2026}				

\usepackage{booktabs}						
\usepackage{multirow}						
\usepackage{amsfonts}						
\usepackage{graphicx}						
\usepackage{duckuments}						
\usepackage{float}

\usepackage[numbers,compress,sort]{natbib}	

\title[Inductive Graph Layout with Implicit Neural Fields]{Inductive Graph Layout with Implicit Neural Fields}

\author[Inal et al.]{%
Berfin Inal\\
Wageningen Univeristy \& Research\\
\email{berfin.inal@wur.nl}\And
Daniel Probst\\
Wageningen Univeristy \& Research\\
\email{daniel.probst@wur.nl}
}

\begin{document}

\maketitle

\begin{abstract}
A graph layout is normally a table of $N$ free coordinates. We optimise a function with a fixed number of parameters instead. This gives a drawing a sample complexity and an extensible domain. Force-directed algorithms remain the standard tools for graph drawing. The most accurate among them minimise stress in the Kamada–Kawai formulation by directly optimising the node coordinates, at a full objective cost of $O(N^2)$ in time and space. Here, we propose Fling (Field Layout via Implicit Neural Geometry), a small neural network mapping the distances of each node to a set of landmarks, positioning it in the plane by training on the layout energy. The full spring system then becomes tractable without its distance matrix, as rest lengths follow from a landmark bound in constant time per pair while a second network learns the majorisation sums from exact anchor rows, at $O(|\mathcal{A}|N)$ per step for $|\mathcal{A}|\ll N$ anchors. Unlike neural drawers that read the graph by message passing, we represent the drawing as a function of node features. An unseen node costs one forward pass, where sparse and low-rank majorisation remain transductive. As the unknowns are weights rather than coordinates, the energy only requires a small fraction of the nodes, and a field fitted that way outperforms PivotMDS, landmark MDS, and a kernel ridge trained on the same energy and features, when the task is fitting the energy of a graph from a sample of its nodes. In addition, the same parameterisation enables a stochastic pivot stress variant, an aesthetics-optimised variant carrying a neighbour-embedding energy with node-edge clearance and crossing terms on the same field, and conditioning on the weight between two energies gives a whole layout family from one run.
\end{abstract}

\section{Introduction}
\label{sec:introduction}

A graph layout gives each node $v \in V$ a position $x_v \in \mathbb{R}^2$ in order to create a readable drawing, where adjacent nodes are close by, edge crossings are minimised, and the graph as a whole spreads out to show its structure. Force-directed methods cast this as energy minimisation, with attraction along edges and repulsion between all pairs of nodes. This includes methods from Fruchterman--Reingold to ForceAtlas2, as well as stress majorisation approaches~\citep{fruchterman1991graph,jacomy2014forceatlas2,gansner2004graph,gansner2012maxent}. All define node coordinates as unknowns in a table $X\in \mathbb{R}^{N\times 2}$, with a free variable per node and dimension. While this is flexible, it comes with two costs: (1) the layout only describes the graph it was computed on. Should a new node be added, it has to be positioned by rerunning the optimiser, as there is no way to map a node to its position. (2) Repulsion is a sum over all $\binom{N}{2}$ pairs, leading to a $O(N^2)$ cost commonly approximated by Barnes--Hut or grid-interpolation schemes.

The strongest available methods solve the stress objective of Kamada-Kawai, who defined a spring between every pair of nodes, with a rest length equal to the graph distance\citep{kamada1989algorithm}. While stress majorisation is the classical solver for Kamada--Kawai, stochastic gradient descent (sgd2) has been introduced to handle the $O(N^2)$ cost of the objective by sampling pairs~\citep{zheng2018graph}. Other approaches include sparse stress, which restricts the pairs to pivots, and MARS, which replaces the distance matrix with a low-rank approximation~\citep{khoury2012drawing,ortmann2016sparse}. As all of these methods optimise free coordinates in the table $X\in \mathbb{R}^{N\times 2}$, they cannot place a node that has not been optimised as part of the graph.

\begin{figure}[h]
\centering
\includegraphics[width=1.0\textwidth]{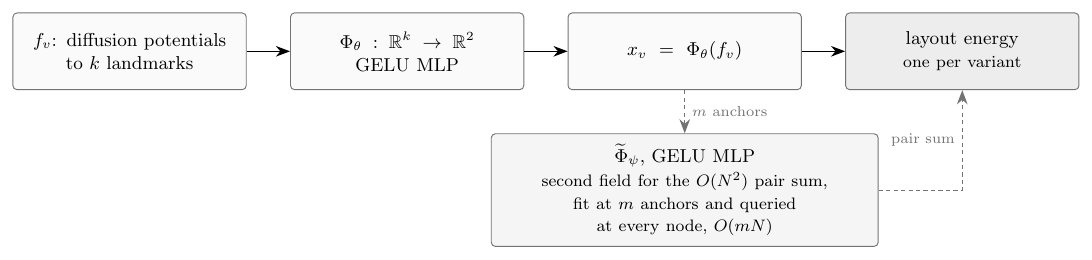}
\caption{The Fling pipeline. All variants read the diffusion potentials $f_v$ as input and drive a coordinate network $\Phi_\theta$ to node positions. They only differ in the energy the weights are trained on, which is all-pairs stress majorisation for Fling, scale-normalised pivot stress for FlingStress, and a neighbour-embedding energy with sampled negatives plus node-edge and crossing terms for FlingVis. The second field $\Phi_\psi$ replaces the $O(N^2)$ pair sum in Fling. It is a training time device, and the forward pass remains $\Phi_\theta$ only.}
\label{fig:arch}
\end{figure}

NeuLay was the first method to move away from the coordinate table by reparameterising the positions as the output of a neural network trained on the same force-directed energy~\citep{Both2023Mar}. As the network learns per-node latents that are optimised together with the network weights, it inherits the problems of previous methods: (1) There is no latent for an unseen node, so it is not able to place one, and (2) it still computes the repulsion over all pairs. Additional learned-layout methods include graph LSTMs that learn the style of a target layout, message-passing GNNs that minimise aesthetic losses on unseen graphs, hierarchical GNNs that can process large graphs, GANs over layout aesthetics, and embedding-plus-stress methods.

Fling (Field Layout via Implicit Neural Geometry) replaces the table with a function. Each node carries a structural feature vector $f_v$, its diffusion potential to a small set of landmarks, computed directly from the adjacency matrix. A coordinate network then maps these features to positions, $p_v = \Phi_\theta(f_v)$ and we optimise the network weights on the layout energy of a single input graph, with majorised stress as the default and pivot-stress and neighbour-embedding energies as the other two variants (Figure~\ref{fig:arch}). Both costs of the table-based approach fall away. (1) The layout is now a map from structure to position, so an added node is placed by one forward pass over its features, with no reoptimisation required, making the drawing inductive. The same holds true for whole graphs drawn from a single population, so a field fitted on a sample places a much larger independent sample by a single forward pass. To write the layout as a function also makes the problem well-posed, as the unknown no longer grows with the graph, so a larger sample refines one map rather than posing a new problem. (2) For the different energies, we solve the $O(N^2)$ pair sum bottleneck in its own way. Fling fits a second, training-time field to the exact sums at $m \ll N$ anchors and query it at every node, at $O(mN)$ per step. The stress and neighbour-embedding variants have sampled objectives, over pivot batches and random negatives, without the need for a second field.

In Table~\ref{tab:neural}, we place Fling (Field Layout via Implicit Neural Geometry) among these other methods. DeepGD, CoRe-GD, and Word2VecGD are pretrained models that learn from a corpus of graphs and then draw an unseen graph via a forward pass~\citep{wang2021deepgd,grotschla2024core,yang2025word2vecgd}. Fling is clearly distinct from such approaches, as it does not require pretraining but instead optimises the weights of a neural network on the layout energy of a single input graph. This makes Fling an alternative to running a force-directed or stress layout on that single graph instead of a model that is trained once and then applied to new inputs. This approach enables out-of-sample (OOS) placement of unseen nodes of a graph it was fit on, rather than drawing a new graph from a learned prior like these corpus-trained methods.

The closest approach to Fling is NNP-Net, which trains per graph and enables intra-graph out-of-sample placement by fitting an MLP on a subgraph and mapping the remaining nodes by inference~\citep{hartskeerl2026nnp}. However, it differs in what the MLP is trained on. NNP-Net initially computes a drawing with a PivotMDS-initialised tsNET variant, which runs t-SNE on graph distances and separates communities that stress layouts draw too close together~\citep{Kruiger2017GraphLB}. It then regresses onto the tsNET coordinates, inheriting the $O(N^2)$ cost and quality ceiling. In comparison, Fling does not require a reference, instead training on the layout energy itself. This allows fling to improve on speed and on the neighbour-embedding energy. A recent multivariate extension to NNP-Net retains the need for a precomputed layout~\citep{hartskeerl2026multivariate}. Another recent approach uses a GNN trained directly on the UMAP objective, amortising across data sets instead of placing nodes into an existing one~\citep{grotschla2026gnn}.

Two further lines of work are relevant to the dense graphs we evaluate Fling on, namely neighbour embedding and local legibility. Neighbour embedding-based graph drawing with tsNET, introduced above, which was made linear in the number of edges through negative sampling by LargeVis and DRGraph~\citep{tang2016vis,Zhu2020DRGraphAE}. Those methods prioritise separation and we take tsNET as the reference to measure our neighbour-embedding results against. What they do not address is local legibility. To solve this, node-edge repulsion methods such as ImPrEd, which is $O(NE)$, were introduced~\citep{Simonetto2011impred}. Recent multicriteria work optimises several such criteria by sampling them under stochastic gradient descent, still using $N$ free coordinates~\citep{ahmed2022multicriteria}.

\section{Methods}
\label{sec:methods}

\subsection{Features, field, and notation}
Fling keeps the layout energies but changes what is optimised. Each node $v \in V$ carries its diffusion potential to the $k$ farthest landmarks, which are a reference set the other nodes are described against, as a structural feature vector $f_v \in \mathbb{R}^k$. Given a random walk with restart probability $\rho$ and length $K$ gives $S=\rho\sum^K_{t=0}(1-\rho)^t\hat{A}^t$, where $\hat{A}=D^{-1/2}AD^{-1/2}$, graph diffusion with a PageRank kernel~\citep{jeh2003scaling}, we take $f_v=\left(-\log(S_{vl}+\epsilon)\right)$ over the landmark set $L$. This is taken as the input to all variants introduced below. As $f_v$ is a deterministic smooth function of $v$'s $K$-hop neighbourhood, any nodes added after training therefore receive features in the same frame as the training nodes and get continuously moved by small changes to the graph, enabling out-of-sample placement without redrawing the complete graph. The layout is then a coordinate network, or implicit neural representation (INR)
$$
x_v=\Phi_\theta(f_v), \Phi_\theta:\mathbb{R}^k\rightarrow\mathbb{R}^2
$$
with an MLP with a GELU activation. The choice of activation function is explained in Appendix~\ref{sec:activation_ablation}. We train the weights $\theta$, not the coordinates, on a layout energy.

Within the Fling framework, we use this parameterisation for three variants that only differ in that energy. (1) Fling minimises all-pairs stress by majorisation, where the rest lengths are supplied by the pivot bound defined in the following paragraph. (2) FlingStress minimises scale-normalised stress against the $|Q|$ exact pivot columns. (3) FlingVis minimises a neighbour-embedding energy jointly with node-edge clearance and crossing terms, and optionally, a distance term. Where a statement refers to all three, we use \textit{the (neural) field}. In some experiments, we refer to the bare coordinate network on one specific energy, we also use \textit{the field}. Furthermore, we reference five specific sets of nodes. (1) Landmarks $L$, where $|L|=k+64$, are selected farthest-first as a frame of reference and give every node the diffusion potential $f_v\in\mathbb{R}^k$. This is the input $\Phi_\theta$ for all three variants. FlingStress additionally read 30 label-keyed probe columns to break ties between structurally equivalent nodes (Appendix~\ref{sec:label_keyed_probes}). Fling and FlingVis read none. We depart from this in experiments where we compare the field against landmark MDS, PivotMDS, and kernel ridge, where every arm uses the same raw standardised BFS distances for comparability. (2) Pivots $P$, where $|P|=16$, are selected farthest-first and the sources of a breadth-first search, yielding hop-distance vectors $h_i\in\mathbb{R}^{|P|}$, from which Fling approximates the rest-length. (3) Target pivots $Q$, where $|Q|=400$, are sampled uniformly and yield the exact BFS distance targets of FlingStress, defined as $\Delta\in\mathcal{R}^{N\times|Q|}$, with $\Delta_{vj}$ the distance from node $v$ to the $j$th pivot. (4) Anchors $\mathcal{A}$, where $|\mathcal{A}|=80$, which are resampled at every step and that are used to evaluate the exact pair term that is used to fit the far field to. From all node sets, only $L$ is used for inference, where a new node reads its $k$ diffusion potentials.

\subsection{Majorised stress without the pair sum (Fling)}
Importantly, no variant computes the pair sum. Only Fling needs a secondary field to avoid doing so. Fling evaluates the exact pair term at $m\ll N$ sampled anchors during each step, fits a secondary field $\widetilde{\Phi}_\psi$ to these values, and queries it at every node. This learned far field does the work that Barnes-Hut does for force sums~\citep{Barnes1986AHO}, without the tree and with $m$ fixed, making each step linear in $N$. In addition, it also rescales the 98th percentile radius at each step, to avoid an inflation of the scale by stray nodes, destabilising anchor dynamics. FlingStress and FlingVis do not require a secondary field, as their objectives are already sampled over pivot batches and random non-adjacent pairs. After training, an unseen node $v^\star$ can be placed by a forward pass $x_{v^\star}=\Phi_\theta(f_{v^\star})$.

Furthermore, Fling admits force laws that free-coordinate methods cannot afford. Barnes-Hut and similar approaches are based on accelerating force sums by pooling distant sources into cells. This works as the same spatial kernel applies to every pair. However, a pair-specific coupling does not admit such an aggregation, resulting in $O(N^2)$ for time and memory. Kamada-Kawai is an example of this and $O(N^2)$ is the reason the objective is, in practice, only used in pair-sampled stress methods. The two reasons for the $O(N^2)$ cost are that the rest lengths need a $N\times N$ distance matrix and the springs cannot be aggregated using a tree. Fling avoids both. We take the $p$ farthest-first pivots and run a BFS per pivot, which gives every node a vector $h_i\in\mathbb{R}^p$ of hop distances at a cost of $O(p(N+E))$. The rest length of any pair is then the pivot bound $r_{ij}=\mathrm{max}_s|h_{is}-h_{js}|$ and evaluated in $O(p)$ per pair. We never calculate a distance matrix. The bound is one-sided (via the triangle inequality $r_{rj}\leq d_{ij}$) and the approximation deciding where Fling works well (Appendix~\ref{sec:pivot_bound}). Importantly, our anchor-field trick isn't specific to the force law we use, as the far field can condition on the per-node vectors exactly as $\Phi_\theta$ does. For majorised stress, it maps the position of a node and 8 of its $p$ hop coordinates to the two components of the majorisation numerator and to the denominator, whose quotient is the Jacobi target. So a pair-specific force costs $O(mN)$.

For Fling, the pair sums are carried via the same trick as the repulsion, namely, a second field that reads the position of the node with eight of its hop coordinates, $\Tilde{\Phi}:\mathbb{R}^{10}\rightarrow\mathbb{R}^3$, where two of the output dimensions are the components of the majorisation numerator and the third is the denominator, whose quotient is the 2D Jacobi target fitted at $m\ll N$ anchors and queried at every node.

\subsection{Distance preservation (FlingStress)}
The majorisation implementation of Fling reaches graph distances through a force law. By training the same coordinate network to reproduce the shortest-path distances with graph distances as an explicit objective, we can match the performance of sgd2 to within a few percent. In order to avoid quadratic complexity, we never calculate the $N\times N$ distance matrix. Instead, FlingStress takes $|Q|\ll N$ uniformly sampled pivots and compute their respective BFS distance values $\Delta\in\mathbb{R}^{N\times q}$ and add the graph edges as additional targets for local fidelity (distance constraints). The energy is computed as 

$$
\mathcal{L}_{ST}=\frac{1}{|\mathcal{T}|+|E|}\left[\sum\limits_{(v,j)\in\mathcal{T}}\left(\frac{ae_{vj}-\Delta_{vj}}{\Delta_{vj}}\right)^{2}+\sum\limits_{(u,w)\in E}\left(ae_{uw}-1\right)^{2}\right]
$$

where $\mathcal{T}=\{(v,j)\in V\times Q:\Delta_{vj}>0\}$ for the pivot pairs at a finite target distance and $x_j$ for the position of the $j$th pivot. The first sum runs over the pivot columns and the second over the edges, whose target length is one hop and

$$
a=\frac{\sum\limits_{(v,j)\in\mathcal{T}}e_{vj}/\Delta_{vj}+\sum\limits_{(u,w)\in E}e_{uw}}{\sum\limits_{(v,j)\in\mathcal{T}}e_{vj}^{2}/\Delta_{vj}^{2}+\sum\limits_{(u,w)\in E}e_{uw}^{2}}
$$

is the scale minimising the bracket in closed form. $a$ is recomputed from current positions with every step and is then held fixed for the duration of that step. The pivot sum is evaluated on a uniformly drawn batch of $\mathrm{min(128,|Q|)}$ columns of $Q$, meaning that the full width of $\Delta$ is never constructed in memory at once. Throughout this paper, this is referred to as the scale-normalised pivot stress. Note that the input and the targets are separate blocks. $\Phi_\theta$ takes the diffusion potentials $f_v$ as input, while the pivot columns $Q$ are targets, never being put through the network. As the network weights are the unknowns, the energy can be summed over a subset of the graph.

\subsection{Neighbour embedding and legibility (FlingVis)}

The third variant, FlingVis, targets neighbour embedding, training the coordinate network on a neighbour-embedding energy with sampled negatives
$$
\mathcal{L}_{NE}=\frac{1}{|E|}\sum\limits_{(u,v)\in E}\log(1+e^2_{uv})-\frac{1}{|\mathcal{N}|}\log\left(\frac{e^2_{uv}}{1+e^2_{uv}}\right)
$$
where $\mathcal{N}$ is a set of uniformly sampled, non-adjacent pairs and the attraction is exaggerated over the first third of training, and adding the node-edge and crossing terms described in the next paragraph. The training runs in two phases, first, the neighbour-embedding energy alone and then the additional terms are added. Each descends a stochastic pair energy for $1\,000$ steps, resulting in the FlingVis being the slowest variant. Instead of Adam, we use Muon as an optimiser (Appendix~\ref{sec:flingvis_optimiser}). Compared to FlingStress, FlingVis reads only the diffusion potentials, with no probes added, as does Fling.

\paragraph{Aesthetic terms of FlingVis.} We showed that we can evaluate a sum on a sample rather than calculating it with the negative sampling. However, this approach is not limited to sums over pairs of nodes. Node-edge and edge-edge repulsion are $O(NE)$ and $O(E^2)$, which is why they are commonly dropped on large graphs. We can keep them by charging each term against a candidate set instead of the full sum. Given a node $v$ and an edge $(u,w)$ that is not incident to $v$, the clearance is the point-to-segment distance
$$
d_{v, uw} = \min_{t \in [0, 1]} \| x_v - (x_u + t(x_w - x_u)) \|,
$$
that is penalised on the violating side only by $\mathrm{relu}(1-d_{v,uw}/c\ell)^2$, where $\ell$ is the median edge length of the current graph drawing and $c$ is the desired clearance factor. Crossings can be reduced to the same type of test, namely, whether the endpoints of each of the segments fall on opposite sides of the other segment's line. $\mathrm{ccw}(u,v,w)$ is the signed area of the triangle, and when setting $d_1=\mathrm{ccw}(p_1,p_2,p_3)$, $d_2=\mathrm{ccw}(p_1,p_2,p_4)$, $d_3=\mathrm{ccw}(p_3,p_4,p_1)$, and $d_4=\mathrm{ccw}(p_3,p_4,p_2)$, two segments cross when $d_1d_2<0$ and $d_3d_4<0$. This means that 
$$
\chi=\sigma\left(-\frac{d_1d_2}{\tau}\right)\sigma\left(-\frac{d_3d_4}{\tau}\right)
$$
where $\sigma$ is the logistic function and $\chi$ the differentiable crossing count. Each $d_i$ is a triangle area twice, meaning that $d_1d_2$ and $d_3d_4$ contains four powers of length. Therefore, we set $\tau=0.25\ell^4$, which makes the argument dimensionless and a product of endpoint clearances in units of a typical edge. The term is only slightly sensitive to the constant $0.25$. A sharper setting only trades a few percent of stress for few crossings. Both terms are only evaluated on the $\kappa$-nearest candidates of $m^\prime$ sampled nodes or edges, costing $O(\kappa m^\prime)$ each step, never forming the full sum.

\paragraph{An optional distance term.} The same objective can also carry distances. We add
$$
\frac{w_{\mathrm{st}}}{|\mathcal{P}|}\sum_{(i,j)\in\mathcal{P}}\left((ae_{ij}-d_{ij})/d_{ij}\right)^{2}
$$
over the sampled pairs $\mathcal{P}$, where $a$ is the scale that minimises the sum in closed form at each step. The term is made useful by refitting $a$ at every term, as holding it fixed would spend the gradient on the size of the drawing, which is not fixed by the neighbour-embedding energy.

\paragraph{Local freedom.} A field has little local, per-node, freedom. As a position is a function of the features, nodes with similar features move together and a clearance term acts on neighbourhoods rather than single nodes. While this coupling is strong, it is not exact (Appendix~\ref{sec:coupling_exactness}). A trainable code per node would remove this constraint, at the price of $dN$ additional parameters and of making the model transductive, costing FlingVis its inductive placement. We chose not to take this route: over nine graphs and three seeds, the trainable code changed the neighbourhood preservation of FlingVis by only $+0.024$ on average.

\subsection{One field for a continuum of objectives}

The input to $\Phi_\theta$ doesn't have to be features only. For a blended energy $\mathcal{L}(\lambda)=(1-\lambda)\mathcal{L}_A+\lambda\mathcal{L}_B$ we supply the network $\Phi_\theta(f_v, \lambda)$ with $\lambda$ as a second input and resample it randomly at every step from a $\mathrm{Beta}(\frac{1}{2},\frac{1}{2})$ prior, which keeps weight on the two endpoints. Conditioning the network on $\lambda$ like this enables us to choose an arbitrary value for it at inference. The naive approach to make the network conditional would be to append $\lambda$ as an additional number to the input so that the network sees $(f_v, \lambda) \in \mathbb{R}^{k+1}$, however, the network tends to underuse $\lambda$ as a scalar among many features (Appendix~\ref{sec:drawing_family}). The result is a family that spans less of the trade-off than one fitted at each weight separately, rather than one that collapses to a single drawing. Instead, we introduce $\lambda$ as Fourier features that modulate the hidden layers through learned per-channel gains and shifts~\citep{Perez2017FiLMVR}. The entire continuum $\lambda \in [0,1]$ is then a single function evaluated at different $\lambda$, where the position of each node across $\lambda$ live in a shared coordinate frame, corresponding directly, with no alignment step. Conditioning a network on the weight of its objective is loss-conditional training introduced by~\citet{Dosovitskiy2020YouOT}.

\section{Experimental setup}

\paragraph{Graphs and protocol.} We chose nine graphs for the benchmark. Meshes and trees, where the pivot bound is tight, and small world graphs where it is not. The graphs are grid\_400 ($N=400$), tree\_400 ($N=400$), delaunay\_600 ($N=600$), sbm\_240 ($N=240$), lesmis ($N=77$), dwt\_1005 ($N=1\,005$), jagmesh8 ($N=1\,141$), 3elt ($N=4\,720$), and ego-Facebook ($N=4\,039$). The meshes are taken from SuiteSparse and ego-Facebook from SNAP. Three additional graphs are used for specific experiments, namely cora ($N=2\,485$, $\mathrm{diag}(G)=19$), and the larger tree and Delaunay variants with $4\,000$ and $1\,000$ nodes, respectively, for scaling and activation experiments. Unless otherwise stated, every number is a mean over three seeds and all Fling experiments run on the GPU, as do the two neural baselines NeuLay and NNP-NET. Stress is the exact all-pairs, scale-normalised stress, computed over every pair.

\section{Results}
Against a full solver, the field costs a few percent of stress and has three properties that a table does not. (1) An energy that runs on a fraction of the nodes, (2) aesthetic terms otherwise dropped for cost, and (3) a drawing that extends to nodes and samples it was never fitted on.

\subsection{How many nodes the energy needs}
As the weights are the unknowns, the energy does not have to run over every node. Table~\ref{tab:main} fits the field of FlingStress on a sample of $M=\mathrm{min}(500,N/2)$ nodes and compares it against baselines that read the same small set of columns and place every node in closed form~\citep{Brandes2006EigensolverMF,de2004sparse}. The choice of $M$ does not decide the ordering. A sweep from $M=50$ to $N/2$ under the same protocol shows that the field leads every other method at every $M$ from $100$ upward, with a widening margin compared to the linear readout. At $M=50$, the closed forms are ahead on one mesh and the linear readout on ego-Facebook (Figure~\ref{fig:sweep}). The closed forms are Landmark MDS, which solves a small block among its landmarks and extends it, and PivotMDS, which scales the full node-by-landmark block directly. The field and the linear readout, which is a single affine layer trained on the same energy, read the same standardised block of BFS columns. The two closed forms, landmark MDS and PivotMDS, read raw columns and are scored at the better of the two blocks available to them, either the 32 landmark columns or the full pivot block. The field, linear readout, and sparse stress choose their learning rate and iteration count on held-out nodes and columns.

\begin{table}[h!]
\centering\footnotesize
\setlength\tabcolsep{4pt}
\caption{Exact all-pairs normalised stress on a sample of $M=\min(500,\,N/2)$ nodes, mean $\pm$ s.d.\ over five seeds.}
\label{tab:main}
\resizebox{\textwidth}{!}{
\begin{tabular}{lrrrrrr}
\toprule
 & & \multicolumn{3}{c}{trained on the sampled energy} & \multicolumn{2}{c}{closed form} \\
\cmidrule(lr){3-5}\cmidrule(lr){6-7}
graph & $M$ & FlingStress & linear & kernel & PivotMDS & landmark MDS \\
\midrule
grid\_400 & 200 & $\mathbf{0.014\,\pm\,0.000}\phantom{^{*}}$ & $0.014\,\pm\,0.001\phantom{^{*}}$ & $0.015\,\pm\,0.001\phantom{^{*}}$ & $0.025\,\pm\,0.001\phantom{^{*}}$ & $0.024\,\pm\,0.001\phantom{^{*}}$ \\
tree\_400 & 200 & $\mathbf{0.043\,\pm\,0.002}^{*}$ & $0.058\,\pm\,0.003\phantom{^{*}}$ & $0.047\,\pm\,0.002\phantom{^{*}}$ & $0.129\,\pm\,0.004\phantom{^{*}}$ & $0.100\,\pm\,0.008\phantom{^{*}}$ \\
sbm\_240 & 120 & $\mathbf{0.144\,\pm\,0.002}^{*}$ & $0.162\,\pm\,0.004\phantom{^{*}}$ & $0.174\,\pm\,0.017\phantom{^{*}}$ & $0.160\,\pm\,0.003\phantom{^{*}}$ & $0.157\,\pm\,0.002\phantom{^{*}}$ \\
lesmis & 38 & $\mathbf{0.114\,\pm\,0.007}\phantom{^{*}}$ & $0.128\,\pm\,0.005\phantom{^{*}}$ & $0.121\,\pm\,0.005\phantom{^{*}}$ & $0.160\,\pm\,0.003\phantom{^{*}}$ & $0.154\,\pm\,0.004\phantom{^{*}}$ \\
delaunay\_600 & 300 & $\mathbf{0.065\,\pm\,0.000}^{*}$ & $0.071\,\pm\,0.001\phantom{^{*}}$ & $0.066\,\pm\,0.001\phantom{^{*}}$ & $0.077\,\pm\,0.001\phantom{^{*}}$ & $0.076\,\pm\,0.001\phantom{^{*}}$ \\
dwt\_1005 & 500 & $\mathbf{0.023\,\pm\,0.000}^{*}$ & $0.029\,\pm\,0.002\phantom{^{*}}$ & $0.025\,\pm\,0.000\phantom{^{*}}$ & $0.029\,\pm\,0.000\phantom{^{*}}$ & $0.029\,\pm\,0.000\phantom{^{*}}$ \\
jagmesh8 & 500 & $\mathbf{0.005\,\pm\,0.000}^{*}$ & $0.009\,\pm\,0.001\phantom{^{*}}$ & $0.006\,\pm\,0.000\phantom{^{*}}$ & $0.016\,\pm\,0.002\phantom{^{*}}$ & $0.010\,\pm\,0.000\phantom{^{*}}$ \\
cora & 500 & $\mathbf{0.108\,\pm\,0.001}^{*}$ & $0.150\,\pm\,0.005\phantom{^{*}}$ & $0.124\,\pm\,0.004\phantom{^{*}}$ & $0.148\,\pm\,0.002\phantom{^{*}}$ & $0.141\,\pm\,0.002\phantom{^{*}}$ \\
tree\_4000 & 500 & $\mathbf{0.040\,\pm\,0.001}^{*}$ & $0.054\,\pm\,0.003\phantom{^{*}}$ & $0.044\,\pm\,0.001\phantom{^{*}}$ & $0.103\,\pm\,0.002\phantom{^{*}}$ & $0.101\,\pm\,0.001\phantom{^{*}}$ \\
3elt & 500 & $\mathbf{0.040\,\pm\,0.000}^{*}$ & $0.052\,\pm\,0.001\phantom{^{*}}$ & $0.042\,\pm\,0.001\phantom{^{*}}$ & $0.060\,\pm\,0.000\phantom{^{*}}$ & $0.057\,\pm\,0.001\phantom{^{*}}$ \\
ego-Facebook & 500 & $\mathbf{0.101\,\pm\,0.002}^{*}$ & $0.114\,\pm\,0.001\phantom{^{*}}$ & $0.132\,\pm\,0.007\phantom{^{*}}$ & $0.231\,\pm\,0.004\phantom{^{*}}$ & $0.219\,\pm\,0.004\phantom{^{*}}$ \\
\bottomrule
\end{tabular}

}
\end{table}

On the mean over seeds, the field is ahead of every matched method on all evaluated graphs. PivotMDS is $1.11$ to $3.19$ times worse (median $\tilde{x}=1.50$) and the landmark MDS $1.10$ to $2.50$ (median $\tilde{x}=1.43$). The linear readout is $1.04$ to $1.77$ times worse with a median of $1.26$, reaching the field only on grid\_400, the graph where the shortest-path metric is a two-dimensional grid and classical scaling is exact. Sparse stress and sgd2 carry free coordinates instead of a map and are references. Sparse stress minimises the same pivot energy over every node, and sgd2 minimises the full all-pair stress.  Table~\ref{tab:lean-controls} replaces the parameterisation with free coordinates. Given the identical sampled energy, a coordinate-table is at or above the random floor on all evaluated graphs.

\paragraph{A smooth interpolant on the same features.} The linear readout as a control only rules out an affine map on the input features, and it may not be necessary to require a network to perform better. To test whether this margin can be closed by any smooth interpolant, we train a kernel ridge regression control on the layout energy itself. We parameterise the positions as $X=\mathcal{K}(F,F_M)W$, where $F$ stacks the features of all node, and $F_m$ those of the $M$ samples. $\mathcal{K}$ is an RBF kernel and $W$ the free coefficients. We optimise $W$ on the same pivot stress over the same samples, and choose its bandwidth, learning rate, and budget by the same rule used by the field. The kernel reaches $1.02$ to $1.30$ times the stress of the field, with a median of $\tilde{x}=1.09$, which is the closest of the matched methods in Table~\ref{tab:main}. The field does not only lead by that margin. The kernel selects a bandwidth on top of the rate and the budget. Placing a node costs the kernel evaluation against all $M$ anchors, whose feature block has to be retained and grows with $M$. In contrast, the $21$k weights and single forward pass of the field do not.

\begin{figure}[h!]
\centering
\includegraphics[width=1.0\textwidth]{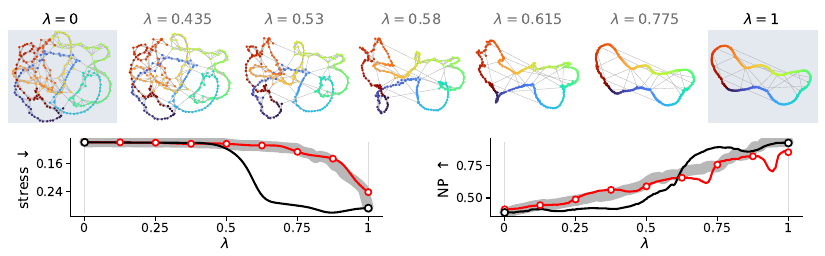}
\caption{The shaded panels show the two $\lambda$ the field was fitted at. The five in between, together with the black line in the line plots, sample at unseen $\lambda$, The grey band represents 60 independent fits and the red line the field resampled at nine $\lambda$ values.}
\label{fig:family}
\end{figure}

\paragraph{A family of drawings from a single run.} Conditioning the field on the weight $\lambda$ between two energies makes the blend between the energies a continuous function, so it can be sampled at any value of $\lambda$. This is similar to INR-enabled superresolution approaches in computer vision~\citep{chen2021learning}. Figure~\ref{fig:family} fits a $N=300$ Watts-Strogatz ring on $\lambda\in\{0,1\}$ alone and then samples it inbetween. Five of the seven panels and 194 further samples, shown as the black line, were never presented to the optimiser, and the knot unties between them, interpolating between a stress- and a neighbour-preservation-optimised drawing. Sixty independent fits, one trained at each $\lambda$ are shown as a wide gray line in the plots. While the independent fits follow the neighbourhood preservation, the true path spends its stress late with only \textasciitilde9\% by $\lambda=0.7$. In comparison, the conditional field moves both at once near $\lambda=0.6$. As the red line in the line plots show, sampling the field along $\lambda$ at only 9 uniformly distributed points largely corrects the interpolation errors.

\subsection{Fling variants against the baselines.} 
\begin{figure}[h!]
\centering
\includegraphics[width=0.9\textwidth]{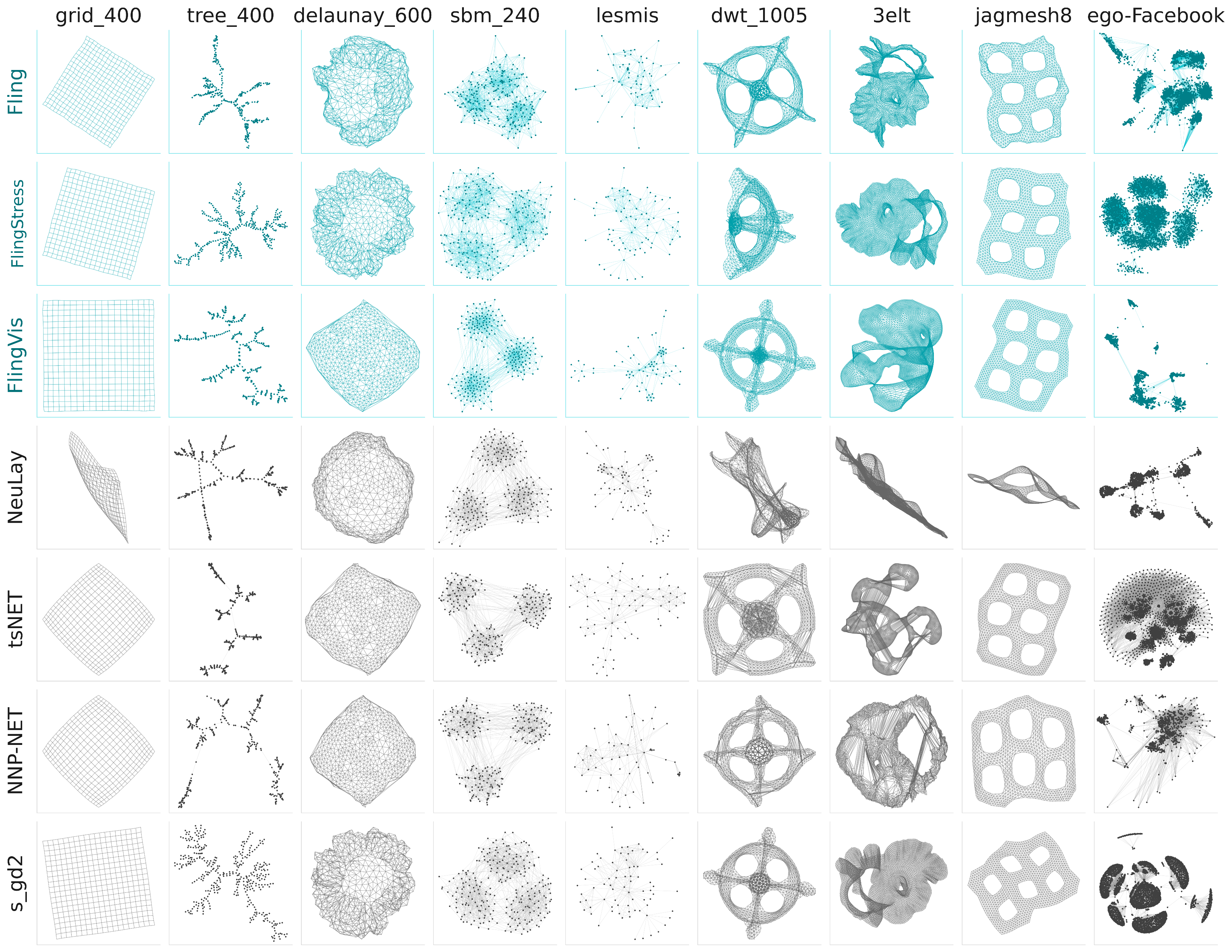}
\caption{Nine benchmark graphs, drawn by the three Fling variants (in colour) and four baselines, one seed per cell. Additional baselines are shown in Figure~\ref{fig:comparison-extra}.}
\label{fig:comparison}
\end{figure}

Figure~\ref{fig:comparison} draws nine benchmark graphs by the three Fling variants and four baselines. Table~\ref{tab:ranks} ranks them, with additional baselines drawn in Figure~\ref{fig:comparison-extra}, by stress, neighbourhood preservation, edge-length uniformity, and crosslessness, averaged over the graphs. 

\textbf{\begin{table}[h!]
\centering\footnotesize
\setlength\tabcolsep{4pt}
\caption{Average rank over $n$ benchmark graphs. Lower is better. Best rank in bold and underlined ranks cannot be separated from the lead (Nemenyi critical difference at $\alpha=0.05$). Crosslessness is not scored on hairballs. sgd2 is excluded to highlight differences between Fling and neural baslines.}
\label{tab:ranks}
\begin{tabular}{lrrrrr}
\toprule
method & overall & stress & neigh.\ pres. & edge-length CoV & crosslessness \\
       &         & $n=9$ & $n=9$ & $n=9$ & $n=5$ \\
\midrule
FlingStress & $3.31$ & $\mathbf{1.44}$ & $\underline{5.00}$ & $\mathbf{1.00}$ & $\underline{5.80}$ \\
FlingVis & $3.41$ & $\underline{4.89}$ & $\mathbf{1.94}$ & $5.22$ & $\mathbf{1.60}$ \\
Fling & $4.44$ & $\underline{2.44}$ & $\underline{5.78}$ & $\underline{2.56}$ & $7.00$ \\
tsNET & $4.55$ & $6.44$ & $\mathbf{1.94}$ & $7.22$ & $\underline{2.60}$ \\
ForceAtlas2 & $5.41$ & $6.00$ & $\underline{4.78}$ & $6.67$ & $\underline{4.20}$ \\
NNP-NET & $5.57$ & $6.44$ & $\underline{3.89}$ & $7.56$ & $\underline{4.40}$ \\
LandmarkMDS & $5.66$ & $\underline{4.56}$ & $7.33$ & $\underline{4.33}$ & $\underline{6.40}$ \\
NeuLay & $6.14$ & $6.89$ & $6.56$ & $5.11$ & $\underline{6.00}$ \\
PivotMDS & $6.50$ & $5.89$ & $7.78$ & $5.33$ & $7.00$ \\
\bottomrule
\end{tabular}
\end{table}}

FlingStress reaches $1.00$ to $1.11$ times sgd2's stress with a median of $1.02$ (Table~\ref{app:tab:stress}). An interesting exception is dwt\_1005, where one of our seeds, which we show in Figure~\ref{fig:comparison}, produces a rare fold that occurs in approximately $6.4$\% of random seeds. Fling falls behind FlingStress on eight of nine graphs, which is the price of training on the pivot bound rather than on the graph distance (Appendix~\ref{sec:pivot_bound}). On ego-Facebook, our variants are at their weakest, due to the small-world deficit of the pivot bound discussed in Appendix~\ref{sec:pivot_bound}.

\begin{figure}[h]
\centering
\includegraphics[width=1.0\textwidth]{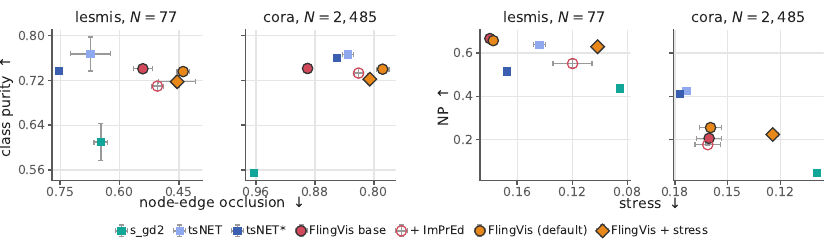}
\caption{The best method is shown at the top right, six seeds, mean $\pm$ s.d. Circles are Fling, squares are baselines. The circle marker shows FlingVis refined by ImPrEd, and the diamond FlingVis with an added stress term at $w_{st}=2$. Occlusion is thresholded at the median edge length}
\label{fig:tradeoff}
\end{figure}

Figure~\ref{fig:tradeoff} scores FlingVis on lesmis and cora against baselines. On both, FlingVis draws the least occluded graph of any method. Adding the two aesthetic terms to a FlingVis variant without them (base) decreases occlusion while keeping purity and stress stable. Introducing the terms as part of the energy also improves upon refining the drawn graph via ImPrEd, which is applied to the FlingVis base variant, at a fraction of the time (430 s and 13 s for ImPrEd and FlingVis, respectively).
Per default, FlingVis does not have a distance term so it trails on the stress column in Table~\ref{tab:ranks}. Adding a distance term at $w_{st}=2$, shown as FlingVis + stress in Figure~\ref{fig:tradeoff}, improves stress with only small cost on neighbourhood preservation and purity,  while occlusion slightly improves.

\section{Conclusion}
We cast graph drawing as fitting a map from node features to positions instead of solving for a table of $N$ free coordinates. The unknowns are a fixed set of weights that are trained on a layout energy of the graph. While this does not necessarily result in a lower minimum, as the comparison of FlingStress with sgd2 shows, it enables an arbitrary choice of energy, inductive placement, and a model size that does not grow with $N$ . Together, they place an unseen node in a single forward pass, draw a whole family of layouts from one fit conditioned on the weight between two energies in one coordinate frame, and optimise clearance and crossings inside the energy rather than refining them afterwards. None of the methods we compare against does all three. The close-form landmark readouts cannot carry a neighbour-embedding or an aesthetic objective at all, while a kernel ridge that can has to keep the features of every anchor, growing with the sample. When fitting the energy of a graph from a sample of its nodes, the field leads all matched readouts. We improve on each learned drawer. NeuLay optimises a per-node latent alongside its weights, meaning that it has nothing to evaluate at a node it never saw. Therefore, it still uses sums over all pairs, leading to wall clock and memory limitations (Figure~\ref{fig:scale}. NNP-Net places unseen nodes as we do, but regresses onto a tsNET reference and inherits its $O(N^2)$ cost and quality ceiling. In comparison, we train on the layout energy itself. The code and data required to reproduce the results in this manuscript can be found in the repository \url{https://github.com/daenuprobst/fling-layout}.

\paragraph{Limitations.} As a position is a function of features, nodes that are not separated by features cannot be separated in the drawing. This is a strong but not exact coupling (Appendix~\ref{sec:coupling_exactness}). To avoid the $O(N^2)$ costs more than sampling noise. Fling never sees the graph distance, as it descends a landmark bound whose majorisation sums are themselves learned. FlingVis charges clearance on nearest-edge candidates instead of all pairs. This means that neither error averages out with more steps. This becomes visible where the bound is loose and Fling trails FlingStress on eight of nine graphs (Appendix~\ref{sec:pivot_bound}. In addition, compared against some methods, the qualitative performance margin is narrow or conditional. A kernel ridge on the same features can reach $1.07\times$ our stress on certain graphs. Finally, comparisons against an affine readout shows that the performance of the network depends on features providing something nonlinear to learn.





\clearpage
\bibliographystyle{unsrtnat}
\bibliography{reference}

\newpage
\appendix
\counterwithin{figure}{section}
\counterwithin{table}{section}

\section{GenAI Usage Statement}
GenAI proofreading tools were used for proofreading and sentence restructuring, as the authors are not native English speakers.

\section{Experimental details}

\subsection{Learned layout methods}

\begin{table}[h]
\centering
\caption{Learned graph-layout methods.}
\label{tab:neural}
\renewcommand{\arraystretch}{1.15}
\setlength{\tabcolsep}{5pt}
\resizebox{\textwidth}{!}{%
\begin{tabular}{@{} l l l l l l r @{}}
\toprule
Method & Model & Trained on & Pretrained & Out-of-sample & Pair sums / cost \\
\midrule
\textsc{Fling} (ours) & coordinate INR & majorised stress / pivot stress / NE & none (per-graph) & new nodes & learned field, $O(mN)$ \\
NeuLay \cite{Both2023Mar} & GCN (adjacency) & FDL energy (wts) & none (per-graph) & no & short-range, $O(N^{2})$ \\
DeepGD~\cite{wang2021deepgd} & MP-GNN & aesthetic loss & corpus (Rome) & new graphs & $O(N^{2})$, OOM ${>}5$k \\
CoRe-GD~\cite{grotschla2024core} & hierarchical GNN & stress & corpus & new graphs & coarsen, sub-quad. \\
Word2VecGD~\cite{yang2025word2vecgd} & node2vec + MLP & cosine-stress & none (per-graph) & no & $O(N^{2})$ stress \\
NNP-Net~\cite{hartskeerl2026nnp} & NN projection & imitate tsNET$^{\star}$ & none (per-graph) & new nodes & sub-quadratic \\
\bottomrule
\end{tabular}%
}
\end{table}

\subsection{Hyperparameters.} 
\label{sec:hyperparameters}
The coordinate network $\Phi_\theta$ is a two-layer GELU MLP with a width of 128 in all experiments, reading $|L|=64$ diffusion landmarks with a restart probability of $\rho=0.05$. $\epsilon$ in $f_v$ is a numerical floor that does not affect the drawing. Fling runs 400 iterations at a learning rate of $5\times10^{-3}$ with $|P|=16$ pivots and $|\mathcal{A}|=80$ anchors, which is the $|\mathcal{A}|$ in $O(|\mathcal{A}|N)$ step-wise cost. FlingStress runs with 500 iterations over $|Q|=400$ pivot columns. FlingVis runs $2\,000$ iterations on the neighbour-embedding energy and $2\,000$ more on the aesthetic terms, weighing each of them with $0.3$ against $4\,096$ sampled negatives. The clearance term uses $c=0.6$ and charges $m^\prime=512$ sampled nodes against their $\kappa=12$ nearest edges, and its crossing term $2\,048$ edge pairs. Counts exceeding a graph are capped, leading to $|Q|$ being reduced to 76 on lesmis and 239 on sbm\_240, $m^\prime$ is reduced to 77 on lesmis. We state deviations where they occur.

\subsection{Metrics}
\label{sec:metrics}
Let $d_{ij}$ be the graph distance between nodes $i$ and $j$ and $e_{ij}=\lVert x_i-x_j\rVert$, which is the distance in the drawing.

Normalised stress as defined by~\citet{Smelser2024NormalizedSI} is written as SNS in the ablation tables and called scale-normalised or exact all-pair stress in the text. It is defined as
$$
\frac{1}{|\mathcal{P}|}\sum_{(i,j)\in\mathcal{P}}\left((ae_{ij}-d_{ij})/d_{ij}\right)^2
$$
where $a$ is the scale minimising it in closed form. It is exact when $\mathcal{P}$ is all $\binom{N}{2}$ pairs. This is what every stress value reports, with the exception of the scaling experiment, and in large graphs where it is sampled where it is declared as such. No variant trains on the exact quantity.

Neighbourhood preservation is the Jaccard overlap between the graph neighbours of a node and its nearest neighbours in the drawing, averaged over nodes. Recall is the same comparison, but scored as the fraction of graph neighbours that were recovered. Edge-length CoV is the standard deviation of drawn edge lengths over their mean. Crlosslessnes, as defined by~\citet{Purchase2002MetricsFG}, is $1-\sqrt{C/C_\mathrm{max}}$, where $C$ is the number of crossing edge pairs and $C_\mathrm{max}$ the number that may cross given the degree sequence. Purity is the fraction of the ten nearest neighbours in the drawing sharing the class of a given node. Occlusion is the fraction of nodes lying closer to a non-incident edge than a quarter of the median edge length. Silhouette is the standard silhouette coefficient of the class labels in the drawing. Ink is the fraction of the drawing covered by edges. Mid is out-of-sample smoothness, namely, the displacement of a held-out node under a small perturbation of its features.

\subsection{Label-keyed probes}
\label{sec:label_keyed_probes}
FlingStress reads 30 random-project columns in addition to the 64 diffusion potentials. They are ten Gaussian probes diffused to three depths, each keyed by splitmix64 on the label of a node rather than on its features. This means it is a pure function of node identity, and adding a node does not influence the features of other nodes. We introduce them to break ties between nodes that are placed at the same point by the diffusion potentials. For example, on dwt\_1005, a mesh with four lobes of the same structure, removing the label-keyed probes FlingStress increases the rates of lobes being folded onto each other from 6 to 18\% over 100 seeds ($p=0.015$), while leaving the stress of the non-folding runs unchanged. At 64 probes, the rate falls to 3\%, meaning that they are solely used for breaking symmetry rather than as a second source of structural information.

\subsection{The pivot bound}
\label{sec:pivot_bound}
Fling trains on the pivot bound $r_{ij}=\mathrm{max}_s|h_{is}-h_{js}|$, over $|P|=16$ farthest-first pivots, rather than the graph distance $d_{ij}$. Triangle inequality gives $r_{ij}\leq d_{ij}$. The bound is exact when a pivot lies beyond one of the endpoints, so that $i$ or $j$ is on a shortest path from $s$ to the other and $|d(s,i)-d(s,j)|=d_{ij}$. It underestimates most when a pivot lies between $i$ and $j$ and the two hop distances cancel each other out. This means that the rest lengths are never too long, only too short. The cost there is that since the majorisation weight is $1/r^2_{ij}$, an underestimated pair is also an overweighted one, leading errors to compound. On a small-world graph, all sixteen pivots can be equidistant from both endpoints, causing the bound to vanish. $r_ij$ is $0$ on 11.7\% of ego-Facebook pairs and none on grid\_400 and 3elt. As the weight $1/r^2_{ij}$ is undefined there, our implementation floors $r_{ij}$ at 0.9 hops, and for these pairs, the floor rather than the bound sets the rest length. The looseness ranks with the deficit of Fling against FlingStress without predicting it.

\subsection{Coupling exactness}
\label{sec:coupling_exactness}
As a position is a function of the features, two nodes sharing the same features would be drawn on top of each. However, even for nodes that the graph cannot distinguish features differ, as they are the diffusion potential to specific landmarks and every node is scored against the same set of labelled landmarks. This means that two interchangeable nodes are still at their own potential from the different landmarks. On a $10\times10$ grid, the transpose and the two reflections are automorphisms that permute the four corners, so no property of the graph distinguishes them. However, their features still differ by up to $7.33$ nats across the landmark columns, and they are drawn at distances from the centre spanning 0.4 to 2.1\% of the mean radius, depending on the variant. That spread is the slack of the coupling. FlingVis, who carries no label-keyed probes, shows the same $7.33$ nats, meaning the slack is caused by the landmarks rather than the probes.

\clearpage
\section{Additional experiments}

\subsection{Full results}

\begin{figure}[h!]
\centering
\includegraphics[width=1.0\textwidth]{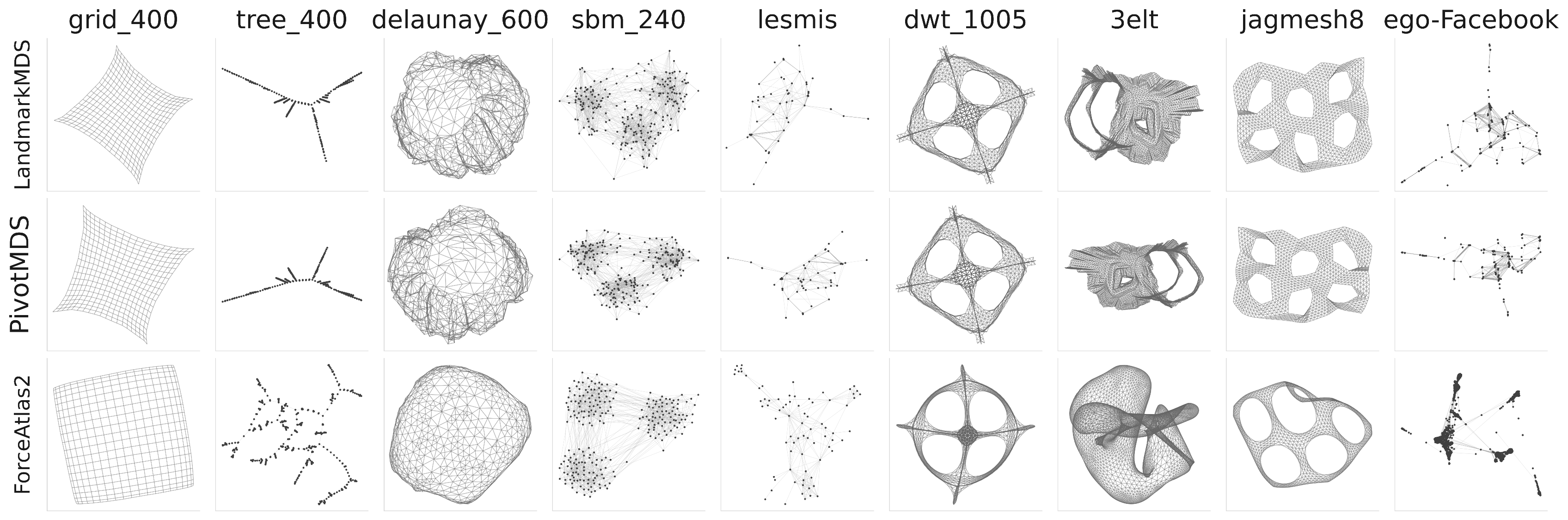}
\caption{The three classical constructions Figure~\ref{fig:comparison} leaves out, drawn the same way: one row per graph, one column per method, single runs at seed $0$.}
\label{fig:comparison-extra}
\end{figure}

\begin{table}[h!]
\centering\footnotesize
\setlength\tabcolsep{4pt}
\caption{Exact all-pairs normalised stress, lower is better. Mean $\pm$ s.d. over three seeds, except FlingVis which is three seeds at $2\,000$ iterations. Bold is best in the row and underline is second best. A dash marks a cell that was not run.}
\label{app:tab:stress}
\resizebox{\textwidth}{!}{%
\begin{tabular}{lcccccccccc}
\toprule
graph & Fling & FlingStress & FlingVis & LandmarkMDS & PivotMDS & s\_gd2 & ForceAtlas2 & tsNET & NNP-NET & NeuLay \\
\midrule
grid\_400 & \underline{0.014}\,$\pm$\,0.000 & \textbf{0.013}\,$\pm$\,0.000 & 0.015\,$\pm$\,0.000 & 0.021\,$\pm$\,0.001 & 0.022\,$\pm$\,0.001 & \textbf{0.013}\,$\pm$\,0.000 & 0.030\,$\pm$\,0.006 & 0.017\,$\pm$\,0.000 & 0.018\,$\pm$\,0.000 & 0.073\,$\pm$\,0.054 \\
tree\_400 & 0.074\,$\pm$\,0.021 & \underline{0.039}\,$\pm$\,0.001 & 0.073\,$\pm$\,0.008 & 0.107\,$\pm$\,0.006 & 0.126\,$\pm$\,0.002 & \textbf{0.036}\,$\pm$\,0.000 & 0.150\,$\pm$\,0.003 & 0.120\,$\pm$\,0.005 & 0.107\,$\pm$\,0.002 & 0.188\,$\pm$\,0.076 \\
delaunay\_600 & 0.070\,$\pm$\,0.003 & \underline{0.064}\,$\pm$\,0.000 & 0.119\,$\pm$\,0.000 & 0.076\,$\pm$\,0.001 & 0.077\,$\pm$\,0.001 & \textbf{0.063}\,$\pm$\,0.000 & 0.097\,$\pm$\,0.000 & 0.119\,$\pm$\,0.000 & 0.118\,$\pm$\,0.000 & 0.121\,$\pm$\,0.067 \\
sbm\_240 & 0.156\,$\pm$\,0.001 & \underline{0.129}\,$\pm$\,0.000 & 0.180\,$\pm$\,0.003 & 0.176\,$\pm$\,0.012 & 0.181\,$\pm$\,0.006 & \textbf{0.128}\,$\pm$\,0.000 & 0.171\,$\pm$\,0.002 & 0.192\,$\pm$\,0.006 & 0.192\,$\pm$\,0.000 & 0.145\,$\pm$\,0.000 \\
lesmis & 0.150\,$\pm$\,0.016 & \textbf{0.086}\,$\pm$\,0.004 & 0.177\,$\pm$\,0.002 & 0.159\,$\pm$\,0.003 & 0.172\,$\pm$\,0.003 & \textbf{0.086}\,$\pm$\,0.003 & 0.137\,$\pm$\,0.009 & 0.143\,$\pm$\,0.005 & 0.173\,$\pm$\,0.002 & \underline{0.129}\,$\pm$\,0.001 \\
dwt\_1005 & \underline{0.024}\,$\pm$\,0.000 & 0.041\,$\pm$\,0.034 & 0.029\,$\pm$\,0.000 & 0.029\,$\pm$\,0.000 & 0.029\,$\pm$\,0.000 & \textbf{0.021}\,$\pm$\,0.000 & 0.065\,$\pm$\,0.042 & 0.055\,$\pm$\,0.001 & 0.049\,$\pm$\,0.001 & 0.162\,$\pm$\,0.040 \\
3elt & 0.042\,$\pm$\,0.001 & \underline{0.039}\,$\pm$\,0.000 & 0.098\,$\pm$\,0.051 & 0.057\,$\pm$\,0.000 & 0.062\,$\pm$\,0.001 & \textbf{0.038}\,$\pm$\,0.000 & 0.161\,$\pm$\,0.010 & 0.165\,$\pm$\,0.008 & 0.122\,$\pm$\,0.003 & 0.239\,$\pm$\,0.047 \\
jagmesh8 & \underline{0.007}\,$\pm$\,0.000 & \textbf{0.005}\,$\pm$\,0.000 & 0.007\,$\pm$\,0.000 & 0.010\,$\pm$\,0.000 & 0.015\,$\pm$\,0.002 & \textbf{0.005}\,$\pm$\,0.000 & 0.118\,$\pm$\,0.056 & 0.009\,$\pm$\,0.000 & 0.017\,$\pm$\,0.000 & 0.156\,$\pm$\,0.024 \\
ego-Facebook & 0.157\,$\pm$\,0.005 & \underline{0.103}\,$\pm$\,0.005 & 0.214\,$\pm$\,0.022 & 0.224\,$\pm$\,0.003 & 0.237\,$\pm$\,0.004 & \textbf{0.093}\,$\pm$\,0.000 & 0.205\,$\pm$\,0.021 & 0.239\,$\pm$\,0.018 & 0.227\,$\pm$\,0.005 & 0.212\,$\pm$\,0.046 \\
\bottomrule
\end{tabular}
}
\end{table}

\begin{table}[h!]
\centering\footnotesize
\setlength\tabcolsep{4pt}
\caption{Neighbourhood preservation, the Jaccard overlap between each node's graph neighbours and its nearest neighbours in the drawing, higher is better. Mean $\pm$ s.d. over three seeds. Bold is best in the row and underline is second best.}
\label{app:tab:neighbourhoodpreservation}
\resizebox{\textwidth}{!}{%
\begin{tabular}{lcccccccccc}
\toprule
graph & Fling & FlingStress & FlingVis & LandmarkMDS & PivotMDS & s\_gd2 & ForceAtlas2 & tsNET & NNP-NET & NeuLay \\
\midrule
grid\_400 & 0.981\,$\pm$\,0.022 & \underline{0.998}\,$\pm$\,0.001 & \textbf{1.000}\,$\pm$\,0.000 & 0.771\,$\pm$\,0.021 & 0.772\,$\pm$\,0.022 & \textbf{1.000}\,$\pm$\,0.000 & 0.758\,$\pm$\,0.054 & \textbf{1.000}\,$\pm$\,0.000 & 0.997\,$\pm$\,0.001 & 0.514\,$\pm$\,0.431 \\
tree\_400 & 0.198\,$\pm$\,0.018 & 0.436\,$\pm$\,0.010 & \textbf{0.626}\,$\pm$\,0.010 & 0.121\,$\pm$\,0.004 & 0.116\,$\pm$\,0.004 & \underline{0.569}\,$\pm$\,0.009 & 0.561\,$\pm$\,0.010 & 0.568\,$\pm$\,0.002 & 0.370\,$\pm$\,0.010 & 0.382\,$\pm$\,0.060 \\
delaunay\_600 & 0.319\,$\pm$\,0.007 & 0.305\,$\pm$\,0.008 & \underline{0.830}\,$\pm$\,0.003 & 0.224\,$\pm$\,0.012 & 0.224\,$\pm$\,0.010 & 0.275\,$\pm$\,0.001 & 0.563\,$\pm$\,0.004 & \textbf{0.834}\,$\pm$\,0.002 & 0.792\,$\pm$\,0.006 & 0.383\,$\pm$\,0.142 \\
sbm\_240 & 0.082\,$\pm$\,0.003 & 0.073\,$\pm$\,0.004 & 0.213\,$\pm$\,0.004 & 0.090\,$\pm$\,0.007 & 0.084\,$\pm$\,0.001 & 0.078\,$\pm$\,0.003 & 0.128\,$\pm$\,0.002 & \textbf{0.264}\,$\pm$\,0.008 & \underline{0.247}\,$\pm$\,0.000 & 0.090\,$\pm$\,0.001 \\
lesmis & 0.375\,$\pm$\,0.072 & 0.423\,$\pm$\,0.009 & \textbf{0.655}\,$\pm$\,0.003 & 0.356\,$\pm$\,0.010 & 0.353\,$\pm$\,0.004 & 0.432\,$\pm$\,0.018 & 0.579\,$\pm$\,0.019 & \underline{0.633}\,$\pm$\,0.020 & 0.497\,$\pm$\,0.005 & 0.555\,$\pm$\,0.025 \\
dwt\_1005 & 0.362\,$\pm$\,0.004 & 0.384\,$\pm$\,0.085 & 0.504\,$\pm$\,0.002 & 0.294\,$\pm$\,0.003 & 0.295\,$\pm$\,0.002 & 0.438\,$\pm$\,0.001 & 0.309\,$\pm$\,0.005 & \textbf{0.572}\,$\pm$\,0.003 & \underline{0.522}\,$\pm$\,0.005 & 0.227\,$\pm$\,0.026 \\
3elt & 0.368\,$\pm$\,0.004 & 0.461\,$\pm$\,0.009 & \underline{0.681}\,$\pm$\,0.158 & 0.274\,$\pm$\,0.002 & 0.272\,$\pm$\,0.002 & 0.440\,$\pm$\,0.000 & 0.327\,$\pm$\,0.013 & \textbf{0.844}\,$\pm$\,0.009 & 0.501\,$\pm$\,0.027 & 0.116\,$\pm$\,0.048 \\
jagmesh8 & 0.677\,$\pm$\,0.004 & \underline{0.729}\,$\pm$\,0.001 & 0.728\,$\pm$\,0.000 & 0.617\,$\pm$\,0.002 & 0.605\,$\pm$\,0.007 & \textbf{0.735}\,$\pm$\,0.000 & 0.375\,$\pm$\,0.054 & 0.721\,$\pm$\,0.000 & 0.684\,$\pm$\,0.023 & 0.263\,$\pm$\,0.061 \\
ego-Facebook & 0.260\,$\pm$\,0.022 & 0.142\,$\pm$\,0.009 & \underline{0.392}\,$\pm$\,0.006 & 0.126\,$\pm$\,0.023 & 0.128\,$\pm$\,0.020 & 0.204\,$\pm$\,0.002 & \textbf{0.437}\,$\pm$\,0.003 & 0.237\,$\pm$\,0.005 & 0.197\,$\pm$\,0.026 & 0.384\,$\pm$\,0.006 \\
\bottomrule
\end{tabular}
}
\end{table}

\begin{table}[h!]
\centering\footnotesize
\setlength\tabcolsep{4pt}
\caption{Edge-length coefficient of variation, the standard deviation of drawn edge length over its mean, lower is better. Mean $\pm$ s.d. over three seeds. Bold is best in the row and underline is second best.}
\label{app:tab:edgecov}
\resizebox{\textwidth}{!}{%
\begin{tabular}{lcccccccccc}
\toprule
graph & Fling & FlingStress & FlingVis & LandmarkMDS & PivotMDS & s\_gd2 & ForceAtlas2 & tsNET & NNP-NET & NeuLay \\
\midrule
grid\_400 & 0.090\,$\pm$\,0.014 & \underline{0.053}\,$\pm$\,0.006 & 0.112\,$\pm$\,0.001 & 0.136\,$\pm$\,0.005 & 0.139\,$\pm$\,0.008 & \textbf{0.014}\,$\pm$\,0.000 & 0.314\,$\pm$\,0.003 & 0.150\,$\pm$\,0.006 & 0.171\,$\pm$\,0.002 & 0.215\,$\pm$\,0.017 \\
tree\_400 & 0.493\,$\pm$\,0.107 & \underline{0.199}\,$\pm$\,0.007 & 0.676\,$\pm$\,0.013 & 0.401\,$\pm$\,0.030 & 0.428\,$\pm$\,0.014 & \textbf{0.176}\,$\pm$\,0.002 & 1.000\,$\pm$\,0.013 & 1.818\,$\pm$\,0.060 & 1.082\,$\pm$\,0.028 & 0.765\,$\pm$\,0.024 \\
delaunay\_600 & 0.384\,$\pm$\,0.014 & \underline{0.319}\,$\pm$\,0.002 & 0.591\,$\pm$\,0.000 & 0.456\,$\pm$\,0.008 & 0.465\,$\pm$\,0.008 & \textbf{0.308}\,$\pm$\,0.001 & 0.551\,$\pm$\,0.001 & 0.633\,$\pm$\,0.003 & 0.595\,$\pm$\,0.002 & 0.453\,$\pm$\,0.047 \\
sbm\_240 & 0.471\,$\pm$\,0.011 & \textbf{0.400}\,$\pm$\,0.002 & 0.894\,$\pm$\,0.015 & 0.529\,$\pm$\,0.012 & 0.570\,$\pm$\,0.004 & \underline{0.413}\,$\pm$\,0.004 & 0.681\,$\pm$\,0.002 & 1.005\,$\pm$\,0.024 & 0.990\,$\pm$\,0.000 & 0.490\,$\pm$\,0.001 \\
lesmis & 0.484\,$\pm$\,0.005 & \textbf{0.358}\,$\pm$\,0.012 & 0.846\,$\pm$\,0.006 & 0.627\,$\pm$\,0.003 & 0.663\,$\pm$\,0.003 & \underline{0.385}\,$\pm$\,0.006 & 0.624\,$\pm$\,0.022 & 0.767\,$\pm$\,0.113 & 0.936\,$\pm$\,0.006 & 0.564\,$\pm$\,0.012 \\
dwt\_1005 & 0.677\,$\pm$\,0.003 & \underline{0.632}\,$\pm$\,0.004 & 0.683\,$\pm$\,0.002 & 0.717\,$\pm$\,0.000 & 0.717\,$\pm$\,0.000 & \textbf{0.622}\,$\pm$\,0.000 & 0.798\,$\pm$\,0.030 & 1.164\,$\pm$\,0.006 & 1.073\,$\pm$\,0.002 & 0.767\,$\pm$\,0.005 \\
3elt & 0.421\,$\pm$\,0.008 & \underline{0.294}\,$\pm$\,0.003 & 0.437\,$\pm$\,0.093 & 0.508\,$\pm$\,0.002 & 0.493\,$\pm$\,0.003 & \textbf{0.278}\,$\pm$\,0.000 & 0.606\,$\pm$\,0.015 & 2.375\,$\pm$\,0.143 & 1.758\,$\pm$\,0.012 & 0.557\,$\pm$\,0.076 \\
jagmesh8 & 0.667\,$\pm$\,0.001 & \underline{0.626}\,$\pm$\,0.002 & 0.630\,$\pm$\,0.000 & 0.667\,$\pm$\,0.000 & 0.670\,$\pm$\,0.002 & \textbf{0.609}\,$\pm$\,0.000 & 0.827\,$\pm$\,0.042 & 0.652\,$\pm$\,0.000 & 0.670\,$\pm$\,0.011 & 0.784\,$\pm$\,0.024 \\
ego-Facebook & 1.022\,$\pm$\,0.030 & \textbf{0.493}\,$\pm$\,0.003 & 2.128\,$\pm$\,0.098 & 2.007\,$\pm$\,0.027 & 2.011\,$\pm$\,0.038 & \underline{0.514}\,$\pm$\,0.000 & 1.523\,$\pm$\,0.038 & 1.226\,$\pm$\,0.039 & 1.833\,$\pm$\,0.171 & 1.085\,$\pm$\,0.006 \\
\bottomrule
\end{tabular}%
}
\end{table}

\begin{table}[h!]
\centering\footnotesize
\setlength\tabcolsep{4pt}
\caption{Every column reports crosslessness, a ratio in $[0,1]$ where higher is better. FlingVis optimises the crossing count itself, so its column is the one to read against the others. Mean $\pm$ s.d. over three seeds, and a dash marks the four graphs whose non-adjacent edge-pair count made crosslessness too expensive.}
\label{app:tab:cross}
\resizebox{\textwidth}{!}{%
\begin{tabular}{lcccccccccc}
\toprule
graph & Fling & FlingStress & FlingVis & LandmarkMDS & PivotMDS & s\_gd2 & ForceAtlas2 & tsNET & NNP-NET & NeuLay \\
\midrule
grid\_400 & \textbf{1.000}\,$\pm$\,0.000 & \textbf{1.000}\,$\pm$\,0.000 & \textbf{1.000}\,$\pm$\,0.000 & \textbf{1.000}\,$\pm$\,0.000 & \textbf{1.000}\,$\pm$\,0.000 & \textbf{1.000}\,$\pm$\,0.000 & \underline{0.999}\,$\pm$\,0.002 & \textbf{1.000}\,$\pm$\,0.000 & \textbf{1.000}\,$\pm$\,0.000 & 0.979\,$\pm$\,0.019 \\
tree\_400 & 0.955\,$\pm$\,0.008 & 0.985\,$\pm$\,0.002 & \textbf{0.996}\,$\pm$\,0.001 & 0.982\,$\pm$\,0.003 & 0.980\,$\pm$\,0.002 & 0.990\,$\pm$\,0.000 & 0.991\,$\pm$\,0.001 & \underline{0.994}\,$\pm$\,0.001 & 0.985\,$\pm$\,0.001 & 0.981\,$\pm$\,0.010 \\
delaunay\_600 & 0.963\,$\pm$\,0.001 & 0.962\,$\pm$\,0.000 & \textbf{0.996}\,$\pm$\,0.001 & 0.953\,$\pm$\,0.002 & 0.953\,$\pm$\,0.001 & 0.959\,$\pm$\,0.000 & 0.979\,$\pm$\,0.000 & \underline{0.990}\,$\pm$\,0.000 & 0.990\,$\pm$\,0.000 & 0.963\,$\pm$\,0.013 \\
sbm\_240 & 0.771\,$\pm$\,0.002 & 0.774\,$\pm$\,0.001 & \textbf{0.814}\,$\pm$\,0.001 & 0.767\,$\pm$\,0.002 & 0.766\,$\pm$\,0.003 & 0.774\,$\pm$\,0.001 & 0.796\,$\pm$\,0.001 & \underline{0.810}\,$\pm$\,0.002 & 0.808\,$\pm$\,0.000 & 0.781\,$\pm$\,0.000 \\
lesmis & 0.798\,$\pm$\,0.030 & 0.812\,$\pm$\,0.007 & \textbf{0.850}\,$\pm$\,0.002 & 0.817\,$\pm$\,0.004 & 0.819\,$\pm$\,0.001 & 0.812\,$\pm$\,0.010 & \underline{0.838}\,$\pm$\,0.002 & 0.838\,$\pm$\,0.013 & 0.807\,$\pm$\,0.001 & 0.830\,$\pm$\,0.006 \\
dwt\_1005 & -- & -- & -- & -- & -- & -- & -- & -- & -- & -- \\
3elt & -- & -- & -- & -- & -- & -- & -- & -- & -- & -- \\
jagmesh8 & -- & -- & -- & -- & -- & -- & -- & -- & -- & -- \\
ego-Facebook & -- & -- & -- & -- & -- & -- & -- & -- & -- & -- \\
\bottomrule
\end{tabular}
}
\end{table}

\subsection{Convergence of the matched-budget comparison}
\label{sec:matched_budget}

\begin{figure}[H]
\centering
\includegraphics[width=1.0\textwidth]{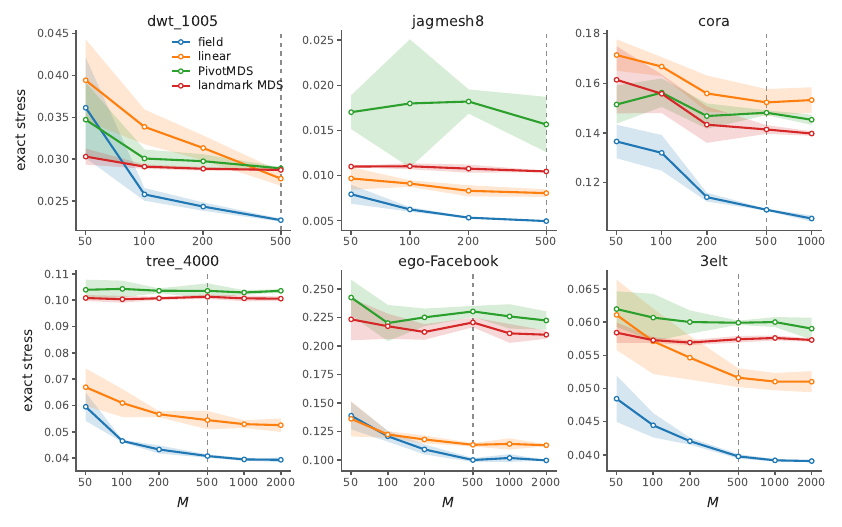}
\caption{Table~\ref{tab:main}'s comparison swept over the sample size $M$, on the six graphs with $N\geq1\,000$, under the identical protocol: the same split rule, the same learning-rate and budget selection on held-out nodes and columns, and the closed forms scored at the better of their two column blocks. Lines are means over three seeds with an s.d.\ band; the dashed line marks the published operating point $M=\min(500,\,N/2)$. The field leads every method at every $M$ from $100$ upward and the published point sits on the flat part of each curve; at $M=50$ the two closed forms are ahead on dwt\_1005 and the linear readout on ego-Facebook.}
\label{fig:sweep}
\end{figure}

\begin{table}[H]
\centering\footnotesize
\setlength\tabcolsep{4pt}
\caption{Exact all-pairs normalised stress for the matched-budget methods. Lower is better. The principal component and the closed forms are matched on information. Mean $\pm$ s.d. over five seeds. Bold is best in row.}
\label{tab:lean}
\resizebox{\textwidth}{!}{
\begin{tabular}{lrrrrrrr}
\toprule
graph & $M$ & FlingStress & linear & kernel & PivotMDS & landmark MDS & PCA (no fit) \\
\midrule
grid\_400 & 200 & $\mathbf{0.014\,\pm\,0.000}$ & $0.014\,\pm\,0.001$ & $0.015\,\pm\,0.001$ & $0.025\,\pm\,0.001$ & $0.024\,\pm\,0.001$ & $0.057\,\pm\,0.012$ \\
tree\_400 & 200 & $\mathbf{0.043\,\pm\,0.002}$ & $0.058\,\pm\,0.003$ & $0.047\,\pm\,0.002$ & $0.129\,\pm\,0.004$ & $0.100\,\pm\,0.008$ & $0.151\,\pm\,0.010$ \\
sbm\_240 & 120 & $\mathbf{0.144\,\pm\,0.002}$ & $0.162\,\pm\,0.004$ & $0.174\,\pm\,0.017$ & $0.160\,\pm\,0.003$ & $0.157\,\pm\,0.002$ & $0.174\,\pm\,0.005$ \\
lesmis & 38 & $\mathbf{0.114\,\pm\,0.007}$ & $0.128\,\pm\,0.005$ & $0.121\,\pm\,0.005$ & $0.160\,\pm\,0.003$ & $0.154\,\pm\,0.004$ & $0.258\,\pm\,0.007$ \\
delaunay\_600 & 300 & $\mathbf{0.065\,\pm\,0.000}$ & $0.071\,\pm\,0.001$ & $0.066\,\pm\,0.001$ & $0.077\,\pm\,0.001$ & $0.076\,\pm\,0.001$ & $0.089\,\pm\,0.002$ \\
dwt\_1005 & 500 & $\mathbf{0.023\,\pm\,0.000}$ & $0.029\,\pm\,0.002$ & $0.025\,\pm\,0.000$ & $0.029\,\pm\,0.000$ & $0.029\,\pm\,0.000$ & $0.068\,\pm\,0.013$ \\
jagmesh8 & 500 & $\mathbf{0.005\,\pm\,0.000}$ & $0.009\,\pm\,0.001$ & $0.006\,\pm\,0.000$ & $0.016\,\pm\,0.002$ & $0.010\,\pm\,0.000$ & $0.069\,\pm\,0.019$ \\
cora & 500 & $\mathbf{0.108\,\pm\,0.001}$ & $0.150\,\pm\,0.005$ & $0.124\,\pm\,0.004$ & $0.148\,\pm\,0.002$ & $0.141\,\pm\,0.002$ & $0.219\,\pm\,0.009$ \\
tree\_4000 & 500 & $\mathbf{0.040\,\pm\,0.001}$ & $0.054\,\pm\,0.003$ & $0.044\,\pm\,0.001$ & $0.103\,\pm\,0.002$ & $0.101\,\pm\,0.001$ & $0.134\,\pm\,0.010$ \\
3elt & 500 & $\mathbf{0.040\,\pm\,0.000}$ & $0.052\,\pm\,0.001$ & $0.042\,\pm\,0.001$ & $0.060\,\pm\,0.000$ & $0.057\,\pm\,0.001$ & $0.096\,\pm\,0.011$ \\
ego-Facebook & 500 & $\mathbf{0.101\,\pm\,0.002}$ & $0.114\,\pm\,0.001$ & $0.132\,\pm\,0.007$ & $0.231\,\pm\,0.004$ & $0.219\,\pm\,0.004$ & $0.236\,\pm\,0.007$ \\
\bottomrule
\end{tabular}

}
\end{table}

\begin{table}[H]
\centering\footnotesize
\setlength\tabcolsep{4pt}
\caption{Floor and references for Table~\ref{tab:lean}. All three methods carry free coordinates and not a map. "table on M rows" gets the sampled energy of the field but has no parameter for the $N-M$ nodes that it has not seen. So it sits above the random floor.}
\label{tab:lean-controls}
\resizebox{\textwidth}{!}{
\begin{tabular}{lrrrrr}
\toprule
graph & $M$ & table on $M$ rows & sparse stress (all $N$) & \textsc{s\_gd2} (all $N$) & random \\
\midrule
grid\_400 & 200 & $0.633\,\pm\,0.034$ & $0.014\,\pm\,0.000$ & $0.013\,\pm\,0.000$ & $0.615\,\pm\,0.001$ \\
tree\_400 & 200 & $0.714\,\pm\,0.005$ & $0.040\,\pm\,0.001$ & $0.036\,\pm\,0.000$ & $0.691\,\pm\,0.002$ \\
sbm\_240 & 120 & $0.367\,\pm\,0.054$ & $0.200\,\pm\,0.022$ & $0.128\,\pm\,0.000$ & $0.295\,\pm\,0.003$ \\
lesmis & 38 & $0.388\,\pm\,0.008$ & $0.124\,\pm\,0.011$ & $0.085\,\pm\,0.002$ & $0.329\,\pm\,0.013$ \\
delaunay\_600 & 300 & $0.497\,\pm\,0.044$ & $0.064\,\pm\,0.001$ & $0.063\,\pm\,0.000$ & $0.495\,\pm\,0.003$ \\
dwt\_1005 & 500 & $0.614\,\pm\,0.046$ & $0.022\,\pm\,0.000$ & $0.021\,\pm\,0.000$ & $0.600\,\pm\,0.001$ \\
jagmesh8 & 500 & $0.729\,\pm\,0.012$ & $0.005\,\pm\,0.000$ & $0.005\,\pm\,0.000$ & $0.675\,\pm\,0.002$ \\
cora & 500 & $0.621\,\pm\,0.004$ & $0.199\,\pm\,0.007$ & $0.099\,\pm\,0.000$ & $0.305\,\pm\,0.001$ \\
tree\_4000 & 500 & $0.913\,\pm\,0.001$ & $0.039\,\pm\,0.001$ & $0.034\,\pm\,0.000$ & $0.781\,\pm\,0.001$ \\
3elt & 500 & $0.867\,\pm\,0.001$ & $0.039\,\pm\,0.000$ & $0.038\,\pm\,0.000$ & $0.641\,\pm\,0.001$ \\
ego-Facebook & 500 & $0.719\,\pm\,0.026$ & $0.201\,\pm\,0.015$ & $0.093\,\pm\,0.000$ & $0.303\,\pm\,0.001$ \\
\bottomrule
\end{tabular}
}
\end{table}

\subsection{Runtime and Memory Scaling}

\begin{figure}[H]
\centering
\includegraphics[width=1.0\textwidth]{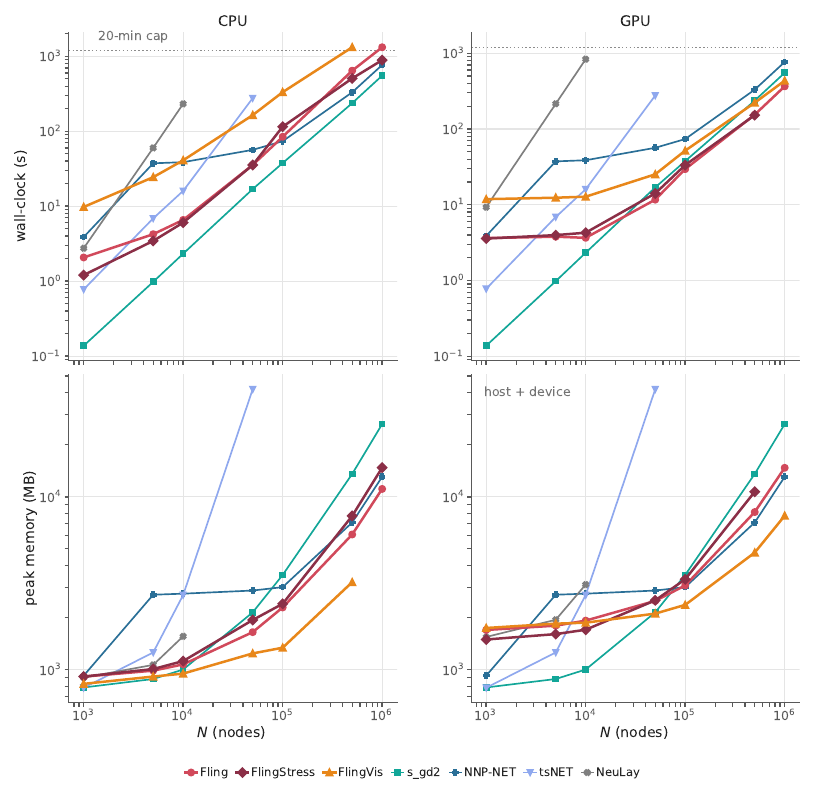}
\caption{Wall clock time and peak memory vs $N$ from $10^3$ to $10^6$ nodes, run on CPU and GPU. Log-log axes. The lines that stop have been gated for extensive wall clock time or memory limits. Single seed at each size.}
\label{fig:scale}
\end{figure}

\subsection{A family of graph drawings from a single run}
\label{sec:drawing_family}
We modulate the hidden layers of a network with $\lambda$ rather than concatenating it to the input. For both variants, the concatenation and the modulation, we read every conditioned field at the two endpoints and compare against unconditioned fields fitted at a single $\lambda$ over nine seeds on sbm\_240. While a bare $\lambda$ column is not ignored, the two endpoints sit $0.196\pm0.014$ apart in RMS node displacement, compared to $0.295\pm0.014$ for the modulated variant. Between $\lambda=0$ and $\lambda=1$, the column variant moves neighbourhood preservation from $0.099$ to $0.158$ and stress from $0.134$ to $0.152$, while modulation moves them from $0.095$ to $0.182$, and $0.132$ to $0.167$, respectively. In comparison, the unconditional fields trained on the respective $\lambda$s span from $0.086$ to $0.200$ and $0.132$ to $0.184$. This means that the column variant reaches about half of the neighbourhood-preservation range and the modulated field about three quarters. The conditional fit draws $n_\lambda$ weights at every step and evaluates the blended energy. This means that one step costs $n_\lambda$ that of the unconditional version.

\clearpage
\section{Ablations}

\subsection{Activation}
\label{sec:activation_ablation}
GELU does not win the ablation outright, however, for each measurement it is within $1.4$\% of the best activation. What is well-separated by the ablation is activation function families. On the two distance-trained variants, the activations tuned for implicit neural representations, namely, SIREN, Gabor, and FINER, leave a field four to seven times less linear between real feature points than plain ReLU or GELU, while costing up to 80\% stress. This means that the standard choice for an INR, for example on images, is not the right choice here. Interestingly, this is not true for FlingVis, where are longer schedule leaves every activation smooth and FINER becomes, in fact, the smoothest.

\begin{table}[H]
\centering
\caption{Activation ablation for the three variants, three-seed mean $\pm$ s.d.\ (per-activation $\omega_0$ as in Table~2). For SNS, lower is better. For recall, higher is better. For mid (out-of-sample smoothness), lower is better. Best per column (by mean) in bold. Delaunay-1000 substitutes the absent \texttt{dwt\_1005}.}
\label{tab:activation_ablation}
\newcommand{\sd}[1]{\,{\scriptsize$\pm$#1}}
\begin{tabular}{@{}lcccc|c@{}}
\toprule
 & \texttt{lesmis} & Delaunay-1000 & \texttt{3elt} & \texttt{cora} & mid $\downarrow$ \\
\midrule
\multicolumn{6}{@{}l}{\texttt{Fling} (majorised stress) (SNS $\downarrow$)} \\
Gabor & $0.152$\sd{0.023} & $0.071$\sd{0.001} & $0.187$\sd{0.001} & $0.310$\sd{0.054} & $1.48$\sd{0.19} \\
SIREN & $0.338$\sd{0.011} & $\mathbf{0.068}$\sd{0.000} & $0.062$\sd{0.030} & $0.239$\sd{0.079} & $1.64$\sd{0.11} \\
ReLU & $\mathbf{0.150}$\sd{0.013} & $\mathbf{0.067}$\sd{0.001} & $\mathbf{0.042}$\sd{0.000} & $\mathbf{0.155}$\sd{0.006} & $\mathbf{0.19}$\sd{0.06} \\
FINER & $0.262$\sd{0.020} & $0.068$\sd{0.001} & $0.045$\sd{0.001} & $0.175$\sd{0.009} & $1.25$\sd{0.42} \\
GELU & $0.151$\sd{0.016} & $\mathbf{0.067}$\sd{0.000} & $\mathbf{0.042}$\sd{0.000} & $0.159$\sd{0.006} & $0.24$\sd{0.09} \\
\midrule
\multicolumn{6}{@{}l}{\texttt{FlingStress} (stress) (SNS $\downarrow$)} \\
Gabor & $0.086$\sd{0.003} & $\mathbf{0.064}$\sd{0.000} & $0.160$\sd{0.064} & $0.216$\sd{0.003} & $1.47$\sd{0.26} \\
SIREN & $0.087$\sd{0.002} & $\mathbf{0.064}$\sd{0.000} & $\mathbf{0.039}$\sd{0.000} & $0.173$\sd{0.009} & $1.17$\sd{0.46} \\
ReLU & $0.087$\sd{0.003} & $\mathbf{0.064}$\sd{0.000} & $\mathbf{0.039}$\sd{0.000} & $0.108$\sd{0.005} & $\mathbf{0.19}$\sd{0.06} \\
FINER & $0.086$\sd{0.002} & $\mathbf{0.064}$\sd{0.000} & $\mathbf{0.039}$\sd{0.000} & $0.110$\sd{0.003} & $0.90$\sd{0.54} \\
GELU & $\mathbf{0.083}$\sd{0.000} & $\mathbf{0.064}$\sd{0.000} & $\mathbf{0.039}$\sd{0.000} & $\mathbf{0.106}$\sd{0.001} & $0.21$\sd{0.07} \\
\midrule
\multicolumn{6}{@{}l}{\texttt{FlingVis} (neighbour embedding) (recall\,\% $\uparrow$)} \\
Gabor & $78.5$\sd{0.8} & $92.0$\sd{0.2} & $87.3$\sd{0.8} & $20.3$\sd{3.5} & $0.53$\sd{0.35} \\
SIREN & $79.4$\sd{0.2} & $92.1$\sd{0.2} & $86.5$\sd{1.0} & $31.5$\sd{1.3} & $0.32$\sd{0.15} \\
ReLU & $\mathbf{79.6}$\sd{0.3} & $\mathbf{92.4}$\sd{0.1} & $81.9$\sd{9.7} & $26.7$\sd{1.4} & $0.21$\sd{0.05} \\
FINER & $79.0$\sd{0.2} & $92.3$\sd{0.0} & $\mathbf{87.7}$\sd{1.0} & $\mathbf{34.2}$\sd{0.9} & $\mathbf{0.18}$\sd{0.05} \\
GELU & $78.9$\sd{0.2} & $92.0$\sd{0.2} & $87.4$\sd{0.6} & $30.9$\sd{5.6} & $0.19$\sd{0.03} \\
\bottomrule
\end{tabular}

\end{table}

\subsection{FlingVis optimiser}
\label{sec:flingvis_optimiser}
Because FlingVis is the slowest variant, we carefully chose the optimiser. Ultimately, we selected Muon to train the coordinate network~\cite{jordan2024muon}. With Muon, each two-dimensional hidden weight is moved along the orthogonal factor of its momentum, which is obtained via five Netwon-Schulz iterations, at $2\times$ the base-rate. Biases and the width-by-two output keep Adam, as this would be just an orthogonalised two-row matrix. For the same reason, per-node code tables keep Adam.

Compared to Adam on the same estimator, Muon reaches 18\% and 10\% higher neighbourhood preservation on cora and ego-Facebook, respectively. In a $500$ to $8\,000$ iteration sweep, Muon matches the best value of Adam in half the steps on cora and in a quarter of the steps for ego-Facebook. This effect is specific to FlingVis and trades minimal distance preservation. 

\subsection{Network Capacity}

\begin{table}[H]
\centering\footnotesize
\setlength\tabcolsep{4pt}
\caption{Width $\times$ depth sweep of the layout network for the default Fling, mean stress relative to the best configuration per graph (grid, sbm, tree, lesmis, jagmesh8, 3elt; two seeds). A flat plateau with one failure mode is large and deep.}
\label{app:tab:capacity}
\begin{tabular}{lcc}
\toprule
$\Phi_\theta$ (width$\times$depth) & rel.\ stress & time \\
\midrule
$256\times2$ & 1.007 & 4.6\,s \\
$128\times3$ & 1.010 & 4.2\,s \\
$128\times2$ (default) & 1.016 & 3.8\,s \\
$64\times3$  & 1.017 & 3.6\,s \\
$64\times2$  & 1.026 & 3.4\,s \\
$256\times1$ & 1.047 & 3.6\,s \\
$128\times1$ & 1.050 & 3.4\,s \\
$64\times1$  & 1.065 & 3.3\,s \\
$256\times3$ & 1.335 & 5.5\,s \\
\bottomrule
\end{tabular}
\end{table}

\end{document}